\documentclass[aps,prl,twocolumn,showpacs,superscriptaddress,floatfix,nofootinbib]{revtex4-1}
\usepackage{graphicx,graphics}
\usepackage{dcolumn}
\usepackage{amsmath,amssymb,amsfonts}
\usepackage{latexsym,verbatim}
\usepackage{bm}
\usepackage{color}
\usepackage[breaklinks=true,colorlinks,citecolor=blue,linkcolor=blue,urlcolor=blue]{hyperref}
\usepackage{tabularx}

\newcommand{\be}{\begin{equation}}
\newcommand{\ee}{\end{equation}}
\newcommand{\bea}{\begin{eqnarray}}
\newcommand{\eea}{\end{eqnarray}}

\begin{document}
\title{Butterflies dragging the jets: chaotic origin of holographic QCD}

\author{Dmitry S. Ageev}
\email{ageev@mi-ras.ru}
\affiliation{Department of Mathematical Methods for Quantum Technologies, Steklov Mathematical Institute of Russian Academy of Sciences, Gubkin str. 8, 119991 Moscow, Russia}

\begin{abstract}

In this Letter, we bring together two topics in the holographic correspondence - quantum chaos and quark-gluon plasma (QGP). We establish that the first relativistic correction to drag force experienced by a charge carrier moving through a thermal medium (for example, a quark in QGP) at a constant velocity is fixed by the butterfly velocity.  Moreover, we show that this result is robust against stringy corrections and anisotropy. For the jet quenching parameter, we find that it is related to the butterfly velocity along the momentum broadening direction and temperature in the spirit of the ``Planckian bound''. This opens a way to the reconstruction of butterfly velocity of quark-gluon plasma and other strongly-coupled systems experimentally from rather simple  observables.
\end{abstract}

\maketitle
Many properties of  strongly-interacting quantum systems are known to carry  large imprint of universality. To explain this universality is a challenging and intriguing problem. A possible way to address it that has successfully been adopted in the past is to search for universal relations between seemingly different and unrelated physical quantities. Holographic correspondence is a versatile and powerful tool to reveal these relations.   The broad range of the holography applications varies from the studies of the heavy-ions collisions and thermal QCD \cite{CasalderreySolana:2011us, DeWolfe:2013cua,Arefeva:2014kyw} to condensed matter theory \cite{Hartnoll:2016apf,ZSL} and quantum information realm  \cite{Rangamani:2016dms}. The discussion about the relation between the  AdS/CFT correspondence  and real physical systems largely started from papers \cite{Kovtun:2004de,Policastro:2001yc} establishing the universal result concerning the viscosity to entropy ratio $\eta/s$ in holographic quantum systems
\be \label{eq:KSSG}
\text{KPSS viscosity relation:}\,\,\,\,\frac{\eta}{s} \sim \frac{1}{4 \pi} \frac{\hbar}{k_{B}}.
\ee 
where ``KPSS'' stands for Kovtun-Son-Starinets-Policatro.
After the experiments at RHIC (for references and review see \cite{CasalderreySolana:2011us}),   many probes amenable to the holographic description have been introduced and studied in the theory of strongly interacting systems and QGP in particular.  
In this work, we focus mainly on the drag force \cite{Herzog:2006gh,Gubser:2006bz} and jet quenching parameter \cite{Liu:2006ug}. Both  are related to the energy and momentum loss for projectiles moving in a strongly interacting quantum system (quark-gluon plasma). However, their origin and properties are slightly different, and we would like  to stress the following points
\begin{itemize}

\item Drag force is associated with the momentum loss of a single quark (charge carrier) moving in the strongly interacting medium. 
\item The jet quenching coefficient $\hat q$ plays the role of the collective transport coefficient (sometimes $\hat q$ is called a jet transport coefficient) describing the momentum broadening in a thermal medium\footnote{The momentum broadening is defined as the probability of momentum increase by the hard parton after propagating through a medium.}. This coefficient is essential for the description of a radiative parton energy loss. It is important to notice that  the jet quenching probes very different scales of medium simultaneously\cite{Casalderrey-Solana:2016iee} .
\item Drag force  also can be related to the transport coefficient, namely conductivity (see \cite{Karch:2007pd}) for the small charge carriers.
\end{itemize}
Recently, the  transport properties of quantum systems have been related to quantum chaos \cite{Blake:2016wvh,Blake:2017qgd}.   
The motivation for this takes its roots in \cite{Hartnoll:2014lpa}, where it was suggested that some velocity $v$ could determine the diffusion constant.
\be \label{eq:Hartnoll}
\text{Hartnoll bound:}\,\,\,\, D \sim \frac{\hbar v^{2}}{k_{B} T}.
\ee 
In \cite{Blake:2016wvh}, this statement has been clarified for   holographic theories with particle-hole symmetry  and  transformed into the relation between charge diffusion\footnote{A similar relation takes place for shear viscosity and diffusion.}, butterfly velocity $v_B$ \cite{Shenker:2013pqa} and the temperature
\be \label{eq:blake}
\text{Blake relation:},\,\,\,\,D_{c}= C\cdot \frac{v_{B}^{2}}{2 \pi T},
\ee 
 where the constant $C$ depends only on the details of the infrared theory, namely $C=d_{\theta}/\Delta_{\chi}$. Here $d_{\theta}$ is the effective spatial dimensionality of the
fixed point  and $\Delta_{\chi}$ is the scaling dimension of the susceptibility. In the paper \cite{Blake:2017qgd}   Blake, Davison and Sachdev (BDS)   extended this relation  to a more general class of theories and thermal diffusion constant $D_T$
\be \label{eq:bds}
\text{BDS relation:}\,\,\,\, D_{T} \sim v_{B}^{2} \tau_{L},
\ee 
where $\tau_{L}=(2 \pi T)^{-1}$ is the Lyapunov time. 

Taking these relations as a prototypical example, we aim to establish a connection between drag force, jet quenching, and butterfly velocity.

$\,$

In this Letter we argue that the drag force and jet quenching parameter also has the imprint of universality analogous to \eqref{eq:KSSG} -\eqref{eq:blake}. We show how they are related to the butterfly velocity and temperature. First of all, we obtain that in the holographic systems   for small velocities $v$ of charge carrier (quark), the properly normalized momentum loss   $dp_\sigma/dt$  is fixed by the butterfly velocity up to a first ``relativistic'', i.e. $v^3$ term
\be \label{eq:dragIntro}
dp_\sigma/dt=-v- {\cal B}\cdot v^3+...,\,\,\,\, {\cal B}= \frac{1}{(d-1) v_B^2},
\ee 
where we denote the normalization factor by $\sigma$.
 One may consider  this identity as the microscopic manifestation of the relations between diffusion and butterfly velocity.
It is worth noticing that,  in principle, this relation allows to measure butterfly velocity in a straightforward manner in experimental setups with charge carriers. Moreover, we provide evidence that this result is robust against higher-derivative corrections on the gravity side and anisotropy.

To reveal a similar universal relation for $\hat q$ one should remember that it is  (especially sensitive) to anisotropy - we always have to specify two directions  (direction of momentum broadening and direction where parton moves). Taking this into account, the relation similar to \eqref{eq:Hartnoll} and \eqref{eq:blake} can also be written down. We argue that the jet quenching can be expressed as
\be \label{eq:intrq}
\hat{q}_y \sim \frac{ T}{\left( v_B^{(y)} \right) ^2 } \sigma
\ee
where $v_B^{(y)}$ is the  butterfly velocity along the momentum broadening direction $y$, $T$ is the temperature and $\sigma$ is the coefficient defining the leading order drag force coefficient acting on the projectiles \eqref{eq:dragIntro}. Another interpretation of $\sigma$ is the ``string tension'' calculated from the asymptotic of spatial Wilson loops \cite{Sin:2006yz}.

The organization of this Letter is as follows. First, we obtain \eqref{eq:dragIntro} and discuss it,  then  turn to the jet quenching parameter and derive relation \eqref{eq:intrq}. In Supplemental Material \cite{supp} we provide all necessary details of calculations.
$\,$

\iffalse
This Letter is organized as follows. In Sec.\ref{sec:drag}
 we discuss and obtain \eqref{eq:dragIntro} and in Sec.\ref{sec:JQ} we derive the relation \eqref{eq:intrq}. In Supplemental Material \cite{supp} we provide all necessary calculation details concerning drag force, jet quenching and butterfly velocity.
 \fi

\section{The chaotic origin of the drag force } \label{sec:drag}
Our main focus is on the $d+1$-dimensional metrics of the form
\be \label{eq:metr}
d s^{2}=-g_{t t} d t^{2}+g_{u u} d u^{2}+g_{ii}  d x^{i} d x^{i},\,\,\,\, i=1,...,d-1,
\ee 
where we assume $g_{ij}$ to be diagonal, and  the horizon  located at $u=u_h$ fixes  
 the temperature and entropy density in dual theory
\be \label{eq:entr}
s=\left.\frac{\sqrt{\det g}}{4 G_{N}}\right|_{u=u_{h}},\,\,\,\,\,T=\left.\frac{\sqrt{\left(g_{t t}\right)^{\prime}\left(g^{u u}\right)^{\prime}}}{4 \pi}\right|_{u=u_{h}},
\ee 
where $G_N$ is a gravitational constant.

$\,$

Consider a heavy particle (quark or charge carrier) moving in the strongly interacting thermal medium with the temperature $T$ at constant velocity $v$. According to holographic duality,  the bulk description of this particle is given  by a classical string hanging from the asymptotic AdS boundary. The particle worldline $x=v\cdot t$ fixes the boundary condition for this string, and this leads us to the string ansatz for the worldsheet $x(t,u)$ of the form
\be 
x(t,u)=v t+\xi(u).
\ee 
As the particle moves through the medium, it experiences momentum loss $dp/dt$ due to the drag force $F=dp/dt$. The calculation of this drag force is well known \cite{Herzog:2006gh,Gubser:2006bz} and the derivation details can be found in  Supplemental Material \cite{supp}. As a result, one can get that the string dynamic depends  on the special bulk point $u_c$ fixed by the condition
\be \label{eq:crpoint}
\left(g_{t t}-g_{x x} v^{2}\right)\Big|_{u=u_{c}}=0.
\ee 
The momentum loss  is defined by $u_c$ as
\be \label{eq:}
\frac{d p_{x}}{d t}=-\frac{v}{2 \pi \alpha^{\prime}}  g_{x x}\Bigg|_{u=u_{c}},
\ee 
where $\alpha^{\prime}$ is the inverse string tension $T_f^{-1}=2\pi\alpha^{\prime}$.
The solution of equation \eqref{eq:crpoint} can be found as a series  \be 
u_c=u_h-\frac{g_{xx}(u_h)}{g_{tt}^{\prime}(u_h)}v^2+...,
\ee 
leading to the expression for the drag force 
\be \label{eq:exp}
F=-\frac{v}{2\pi\alpha^{\prime}} g_{xx}(u_h)-\frac{v^3}{2\pi\alpha^{\prime}} \left(\frac{g_{xx}^{\prime}}{g_{tt}^{\prime}}\cdot g_{xx}\right)\Big|_{u=u_h}+... 
\ee 
Now  turn to the chaotic properties of the holographic dual described by \eqref{eq:metr}.
There are different   characteristics of quantum system relevant to quantum chaos recently proposed to be calculated \cite{larkin,Shenker:2013pqa,Roberts:2014isa, Shenker:2013yza} from  the exponential growth of Hermitian operators commutators
\be 
\left\langle\left[{\cal O}_x\left(t_{w}\right), {\cal O}_{y}(0)\right]^{2}\right\rangle_{\beta} \sim   e^{\lambda_{L}\left(t_{w}-\tau_{*}-|x-y| / v_{B}\right)}.
\ee 
Here scrambling time $\tau_*$  is the time of the chaos onset, butterfly velocity $v_B$ defines the effective lightcone constraining the spatial chaos spreading and Lyapunov exponent $\lambda_L$ is related to the chaotic features of time evolution. In holographic correspondence one can calculate\footnote{See Supplemental Material \cite{supp} for derivation of the butterfly velocity and references.} $\tau_*$, $\lambda_L$ and $v_B$ for the quantum system dual to \eqref{eq:metr} in terms of metric  components values at the horizon $u_h$
\be \label{eq:chaos}
\lambda_L=2 \pi T,\,\,\,\,\,\,v_B^2=\frac{g_{tt}^{\prime}(u_h)}{ g_{xx}^{\prime}(u_h) (d-1)}.
\ee 
Combining \eqref{eq:entr}, \eqref{eq:chaos}, \eqref{eq:exp} and expanding for small $v$  we obtain, that  momentum loss  $dp/dt$ normalized by the leading order coefficient $\sigma$  depends only on the butterfly velocity   at the first sub-leading order in $v$
\be \label{eq:expvb} 
\frac{dp_\sigma}{dt}=\frac{1}{\sigma}\frac{dp}{dt}=-v-v^3 \cdot \frac{1}{(d-1) v_B^2} +...,.
\ee 
where 
\be \label{eq:sigma}
{ \sigma}= (2\pi\alpha^{\prime})^{-1} g_{xx}(u_h),\ee 
and $p_\sigma= p/\sigma$ is the normalized momentum. A few comments are in order now
\begin{itemize}
    \item We propose that using this formula, one can determine the butterfly velocity of the strongly interacting quantum system in the quite general experimental setup and for a wide range of quantum systems where the measurement of momentum loss is possible. The  experimental setups and protocols allowing the measurement of butterfly velocity have been widely discussed previously \cite{Swingle:2016var,Yao:2016ayk,Garttner:2016mqj,Li:2017pbq}. The drag force relatively simple observable and our proposal can be applied to a quite broad range of quantum systems (at least those one with the massive charge carriers or (quasi)particle)

\item The leading order coefficient $\sigma$ can be interpreted as a so-called spatial string tension. This quantity determines the asymptotic behavior of the spatial Wilson loops with large spatial extents. This interpretation has been noticed first in \cite{Sin:2006yz} and studied further in \cite{Andreev:2017bvr,Arefeva:2020vhf,Arefeva:2020bjk}. If we consider gauge theories and the application to QGP, $\sigma$ is responsible for chromomagnetic fluctuations in our medium. However, we would like to stress that one can avoid this interpretation and consider $\sigma$ as the leading order drag force coefficient.

\item An important issue in the identities like \eqref{eq:KSSG} or \eqref{eq:bds} is the robustness of such results against  stringy corrections and anisotropy. It is known that KPSS bound \eqref{eq:KSSG} is violated by stringy corrections and anisotropy while BDS identity is robust. In Supplemental Material \cite{supp} we provide evidence that our proposal is robust against stringy corrections.  It is also robust against the inclusion of anisotropy with a very mild modification. Namely, we derive that the butterfly velocity $v_B^{(i)}$ along the  spatial direction $i$ depends on the first and second drag force coefficients along different direction $F_i=-\sigma_i v-{\cal B}_i v^3$ as
\be 
v_B^{(i)}=\sqrt{\frac{\sigma_i^{-1}}{\sum_{i=1}^{d-1}{\cal B}_i\sigma_i^{-2}}}.
\ee 
\end{itemize}

\section{Jet quenching and butterfly velocity}
\label{sec:JQ}
Another interesting quantity called a {\it jet quenching parameter}  $\hat q$  also characterizes the energy and momentum loss of projectiles in a strongly interacting medium. In the studies of heavy-ions collisions and thermal QCD, the jet-quenching phenomena are called the disappearance or suppression of the bunch of hadrons resulting from the fragmentation of a  parton after strong interaction leading to momentum loss in the dense medium (quark-gluon plasma). In general, the details of the mechanism of a jet energy loss depend on the medium properties. The jet quenching parameter can be defined in the perturbative framework and considered as a kind of transport coefficient. Also, it allows a non-perturbative definition in terms of adjoint light-like Wilson loop \cite{Liu:2006ug}   useful in the gauge/gravity duality 
\be 
\left\langle W^{A}(\mathcal{C})\right\rangle \approx \exp \left[-\frac{1}{4 \sqrt{2}} \hat{q} L^{-} L^{2}\right].
\ee 
Here $L_-$ corresponds to the distance between  the light-like parts of the contour $\mathcal{C}$ and $L$ between the transversal one with $L_{-} \gg L$.  From the holographic viewpoint, this Wilson loop can be calculated by the string hanging from the light-like contour on the boundary, and we leave all details of calculations in Supplemental Material \cite{supp}. 

$\,$

In general, we have in mind a dual metric of the form \eqref{eq:metr} and in particular, the geometries similar to the background with hyperscaling violation to make parallels with \cite{Blake:2016wvh}. However, there are some possible subtleties related to the divergences stemming from UV structure of this kind of theory. It is worth stressing that while the stringy jet-quenching calculation is typically considered for a five-dimensional gravitational background, we assume an arbitrary space-time dimension.  Also, we do not discuss the gravitational action that gives our metric  as a solution and just takes quite a general form of ansatz because the jet-quenching formula depends only on metric details. Of course, one could meet some restrictions on the metric coefficients like in the hyperscaling violating theories. 

$\,$

We restrict our attention to $d$-dimensional metric \eqref{eq:anis} of the form
\be \label{eq:anis} 
ds^2=-\frac{f(z)}{z^{2\nu}}dt^2+\frac{dz^2}{z^{2\nu_z}f(z)}+\frac{dx^2}{z^{2\nu}}+\frac{dy^2}{z^{2\nu_y}}+\sum_{\alpha}\frac{dx_\alpha^2}{z^{2\nu_\alpha}} ,  
\ee 
 and  assume that the  parton is moving along direction $x$ while the momentum broadening occurs along $y$. 
  We refer the reader
to Supplemental Material \cite{supp}  for the computational details and derivation of the formula for jet quenching parameter   \cite{Giataganas:2012zy}\footnote{Notice the sign in the definition of $g_{tt}$ component.} which has the form
\be \label{eq:jqintegral}
\hat{q}_y=\frac{\sqrt{2}}{\pi \alpha^{\prime}}\left(\int_{0}^{u_{h}} \frac{1}{g_{yy}} \sqrt{\frac{g_{u u}}{g_{--}}}du\right)^{-1}, \,\,\,\,g_{--}=\frac{g_{xx}-g_{tt}}{2},
\ee 
for a general class of anisotropic metrics  \eqref{eq:metr}.
The jet quenching parameter temperature dependence corresponding to the isotropic background with $\nu=\nu_t=\nu_z=\nu_y=\nu_{\alpha}=1$ derived first in \cite{Liu:2006ug} has the form
\be 
q= {\cal B}_0 \cdot T^3,
\ee 
where ${\cal B}_0$ is some constant. The butterfly velocity for this choice of parameters is temperature-independent. Thus it is not clear whether the jet quenching is related to it.  
One should notice that from the very beginning, jet quenching is intrinsically sensitive to anisotropy (see \cite{Giataganas:2012zy,Chernicoff:2012gu,Rebhan:2012bw,Ageev:2016gtl,Ageev:2016viy} for holographic studies of jet quenching parameter in the presence of anisotropy). We have to specify two directions for jet quenching in contrast to the drag force and conductivity.  To reveal some non-trivial relation between $\hat q$ and chaotic characteristics, we focus on the jet quenching for some particular but still quite general anisotropic metric with $\nu_t=\nu_x=\nu$. We consider two  different butterfly velocities $v_B^{(x)}$ and $v_B^{(y)}$ associated with  spatial directions $x$ and $y$ respectively\footnote{A brief review concerning the derivation of the anisotropic butterfly velocities can be found in Supplemental Material \cite{supp}}.
For metric \eqref{eq:anis} they have the form 
\be \label{eq:vxvy}
v_B^{(x)}=\sqrt{\frac{a}{2(\sum_i \nu_i)}}  ,\,\,\,\,\,v_B^{(y)}=\sqrt{\frac{a}{2(\sum_i \nu_i)}} z_h^{ \nu_y-\nu},
\ee 
where $\sum_i$ is the summation over all spatial $\nu_{i}$.
For our choice of anisotropic exponents the jet quenching parameter depends on $z_h$ as
\be \label{eq:jqzh}
q= \frac{1}{2\pi\alpha^{\prime}}{\cal B}\cdot z_h^{-2 \nu_y+\nu_z-\nu-1},
\ee 
where  ${\cal B}$ depends only on exponents $\nu_{x}$, $\nu$, $\nu_z$ and spacetime dimension $a$ (see Supplemental Material \cite{supp} for the explicit form of ${\cal B}$).
We are looking for the relation between the jet quenching parameter and butterfly velocity supplemented with the additional characteristic depending on the inverse string tension. The equation \eqref{eq:jqzh} being combined with the temperature of the metric \eqref{eq:anis} 
\be \label{eq:entr4}
T=\frac{a z_h^{\nu_z-\nu-1}}{4 \pi },
\ee 
and butterfly velocity $v_B^{(y)}$ defined by \eqref{eq:vxvy}
results in the relation
\be \label{eq:2djq}
q_y = {\cal A} \cdot \frac{ T}{\left( v^{(y)}_B \right)^2} \sigma,
\ee 
between the jet quenching parameter,   the square of butterfly velocity $v_B^{(y)}$,  and the leading order drag force coefficient \eqref{eq:sigma}. The constant ${\cal A}$  depends only on the dimension $a$ and the infrared exponents $\nu$ as it should be if we are looking for the analog of \eqref{eq:blake} or \eqref{eq:bds}. Again one can interpret $\sigma$ as the string tension calculated from the asymptotic behavior of the spatial Wilson loop. However, it seems more natural here to consider $\sigma$ in terms of the drag force acting on the projectiles.

As we mentioned before, the jet quenching probes very different scales of the system: an initial fragmentation of parton which is weakly-coupled and late-time interaction of jets with a thermal medium which needs non-perturbative description \cite{Casalderrey-Solana:2016iee}. The relation \eqref{eq:2djq} involves the thermodynamics (temperature $T$) and late time drag force coefficient $\sigma$, which defines the dynamics of slowly moving projectile. As we have shown before on the intermediate non-perturbative scales, the  drag force acting on the projectile is described by the butterfly velocity which is also present in \eqref{eq:2djq}.

$\,$

\section{Discussion}
In summary, we have obtained two relations between probes  in strongly coupled quantum theory and butterfly velocity $v_B$.   Both results are obtained in the framework of holographic correspondence. Firstly, we have shown that the sub-leading (i.e. first ``relativistic'') coefficient in the drag force is fixed by butterfly velocity $v_B$. This leads us to the possibility of measuring butterfly velocity by experimental study of the velocity dependence for momentum loss of charge carriers in strongly coupled theories. Secondly, analogous to the charge diffusion constant \cite{Blake:2016wvh}, the jet quenching coefficient is defined by anisotropic butterfly velocity, temperature, and leading order drag force coefficient up to some constant.  The jet quenching results are obtained for quite general theory dual to a metric similar to anisotropic hyperscaling geometry.  The presented results indicate that these quantities in strongly coupled theories and, particularly, quark-gluon plasma are governed by butterfly velocity and thermodynamic quantities.

Let us briefly discuss possible future extensions of this work 
\begin{itemize}
    \item It would be interesting to extend the understanding of the relation between drag force  and butterfly velocity in different directions, such as the corrections caused by quark mass or other drag force proposals \cite{Ficnar:2013qxa} (for review of quark dynamic holographic description see \cite{Chernicoff:2011xv}).  Also, it is interesting to consider the anisotropic background where particle moves in an arbitrary direction and study a similar relation for the butterfly velocity and Wilson loops \cite{Arefeva:2020vhf,Arefeva:2020bjk,Arefeva:2021jpa}.
    \item Another prospective question is to consider the phenomenological implications of our identities using the known backgrounds reproducing  experimental results concerning drag force and jet quenching (see for example \cite{Ficnar:2013qxa,Gursoy:2009kk}).
    \item Finally, the intriguing direction to study is  the relation of chaos to other  probes of QGP, including hot wind \cite{Liu:2006nn}, glueball spectrum \cite{Karch:2006pv} or particles production multiplicity \cite{Gubser:2008pc,Arefeva:2009pxq}. An interesting proposal revealing some relation between Lyapunov exponent and QCD  recently appeared \cite{deBoer:2017xdk,Akutagawa:2019awh,Colangelo:2020tpr,Liu:2018gae}, and it would be very interesting to find the connection between our proposal and described in these papers.
\end{itemize}

\end{document}

\title{Supplemental material for ``Deterministic chaos and fractal entropy scaling in 2d Floquet CFT''}
\author{Dmitry S. Ageev}
\author{Andrey A. Bagrov}
\author{Askar A. Iliasov}

\maketitle
\section{Drag force and jet quenching} 
\subsection{Drag force calculation}
\label{sec:dragjq}
We consider a (quasi)particle moving with constant velocity $v$ through some strongly interacting quantum system and experiencing the drag force due to interaction with it. A quark moving through a strongly interacting quark-gluon plasma formed in the heavy-ion collision was the first model of this type described in holography \cite{Herzog:2006gh,Gubser:2006bz}.  
In the holographic correspondence, one typically considers  a quantum system at finite temperature or finite chemical potential, which is  dual to the background with horizon and  
$d+1$-dimensional metric of the form
\be \label{eq:metr-app}
d s^{2}=-g_{t t} d t^{2}+g_{u u} d u^{2}+g_{ii}  d x^{i} d x^{i},\,\,\,\, i=1,...d-1
\ee 
where
\be 
g_{t t} \sim c_{0}\left(u-u_{h}\right)\Big|_{u\rightarrow u_h}, \quad g_{u u}=\frac{c_{1}}{\left(u-u_{h}\right)}\Big|_{u\rightarrow u_h}.
\ee 
A holographic dual of the particle (quark) is the classical string and the ansatz for it is taken in the form
\be 
x(t,u)=v t+\xi(u),\,\,\,\, \xi(u)\Big|_{u\rightarrow u_{bnd}}=0,
\ee 
such that at the asymptotic boundary $u\rightarrow u_{bnd}$  the string swipes the particle worldline  $x=v\cdot t$. 
The Nambu-Goto action for this ansatz is
\be 
S=-\frac{1}{2 \pi \alpha^{\prime}} \int \mathcal{L} \, d \tau d\sigma   ,\,\,\,\,\mathcal{L} =  \sqrt{-\operatorname{\det} g},
\ee 
and it can be written down explicitly in terms of metric components
\be \nn
\mathcal{L} =  \sqrt{-\operatorname{det} g}= \sqrt{-g_{uu} g_{t t}-g_{uu} g_{x x} v^{2}-g_{x x} g_{t t} \cdot \xi^{\prime 2}(u)},
\ee 
where we use the gauge $\sigma=u, \tau=t$. 
The equations of motion for $\xi(u)$ imply that
\be 
\pi_{X}=\frac{\partial \mathcal{L}}{\partial \xi^{\prime}}= \frac{g_{x x} g_{t t}}{\sqrt{-g}} \xi^{\prime}
\ee 
is a constant. The solution\footnote{In this solution $\pi_{\xi}$ is some constant unimportant for our purpose.} of this equation with respect to $\xi^{\prime}$ has the form
\be \label{eq:app-xipr}
\left(\xi^{\prime}\right)^{2}=-\pi_{\xi}^{2} \frac{g_{uu}\left(g_{t t}-g_{x x} v^{2}\right)}{g_{x x} g_{t t}\left( g_{x x} g_{t t}+\pi_{X}^{2}\right)},
\ee 
and the special bulk point $u_c$ defined by
\be \label{eq:app-crpoint}
\left(g_{t t}-g_{x x} v^{2}\right)\Big|_{u=u_{c}}=0,
\ee
the numerator in \eqref{eq:app-xipr}  changes the sign. This condition fixes
\be 
\pi_{X}^{2}=-\left. g_{x x} g_{t t}\right|_{u=u_{c}}=\left.v^{2}  g_{x x}^{2}\right|_{u=u_{c}},
\ee 
where  we used \eqref{eq:app-crpoint}. The   current density for momentum along direction $x$ is given by
\be 
P_{x} =-\frac{1}{2 \pi \alpha^{\prime}}  g_{x \nu} g^{u \beta} \partial_{\beta} X^{\nu} 
=-\frac{1}{2 \pi \alpha^{\prime}}  \frac{g_{x x} g_{t t}}{\det g} \xi^{\prime} ,
\ee 
and it is related to the quark momentum loss as
\be 
\frac{d p_x}{d t}=\sqrt{-\det g} P_{x}=-\frac{1}{2 \pi \alpha^{\prime}}  \frac{g_{x x} g_{t t}}{\sqrt{\det g}} \xi^{\prime} ,
\ee 
which after some algebra simplifies to
\be
\frac{d p_x}{d t}=-\frac{\pi_{X}}{2 \pi \alpha^{\prime}}=-\left.\frac{v}{2 \pi \alpha^{\prime}} g_{x x}\right|_{u=u_{c}}.
\ee 
Typically the equation \eqref{eq:app-crpoint} cannot be solved exactly, and the solution can be found as a series in $v$. However, there are some important cases when the exact solution can be found explicitly. The simplest well-known answer for the drag-force temperature and velocity dependence corresponds to the $AdS$-black hole solutions in $d+1$-dimensional Einstein gravity with the metric
\begin{gather} \label{eq:einst}
f(u)=1-u^d/u_h^d,  \\
ds^2=\frac{1}{u^2}\left(-f(u)dt^2+\frac{du^2}{f(u)}+dx^2 \right),
\end{gather}
with the exact solution for drag force
\be 
F=-\frac{1}{2\pi\alpha^{\prime}}\frac{16 \pi ^2 T^2}{d^2}\frac{v}{\left(1-v^2\right)^{2/d}}.
\ee 
Expanding up to the sub-leading order in $v$ and normalizing on the coefficient in front of the leading order term
\be 
F\approx-v-\frac{2}{d}v^3
\ee 
we find, that according to our main proposal the subleading coefficient is given by the inverse square root of butterfly velocity $v_B$ for dual of \eqref{eq:einst}, where
\be 
\frac{2}{d}=\frac{1}{v_B^2}\frac{1}{(d-1)}\,\,\,\,\,\,v_B=\sqrt{\frac{d}{2(d-1)}}.
\ee 
It is interesting to take a look at what happens if we include a higher-derivative correction to a background \eqref{eq:einst}. On the boundary side this corresponds to the inverse 't Hooft coupling corrections. For simplicity,  let us focus on the five-dimensional background considered explicitly in \cite{Fadafan:2008gb,VazquezPoritz:2008nw}.
 The $AdS$ black hole with stringy corrections included  is the solution of Gauss-Bonnet gravity \cite{Cai:2001dz} and has the form 
\be
ds^{2}=-N^{2} \frac{r^{2}}{R^{2}} f(r) d t^{2}+\frac{d r^{2}}{\frac{r^{2}}{R^{2}} f(r)}+\frac{r^{2}}{R^{2}} d \vec{x}^{2}
\ee
where
\be 
f(r)=\frac{1}{2 \lambda_{\mathrm{GB}}}\left(1-\sqrt{1-4 \lambda_{\mathrm{GB}}\left(1-\frac{m R^{2}}{r^{4}}\right)}\right), \ee 
and 
\be 
N^{2}=\frac{1}{2}\left(1+\sqrt{1-4 \lambda_{\mathrm{GB}}}\right)
\ee
After straightforward calculation \cite{Fadafan:2008gb,VazquezPoritz:2008nw,Ficnar:2013qxa}, we obtain that the drag force for this case is given by
\be \label{eq:FGB}
F\left(\lambda_{\mathrm{GB}}\right)=-\left(\frac{\pi \sqrt{\lambda}}{2} T_{\mathrm{GB}}^{2}\right) \frac{v}{\sqrt{N^{4}-N^{2} v^{2}+\lambda_{\mathrm{GB}} v^{4}}}.
\ee 
where $\lambda$ is 't Hooft coupling and
\be 
T_{\mathrm{GB}}=\frac{N r_{h}}{\pi R^{2}}.
\ee 
 Expanding \eqref{eq:FGB} we obtain the expression for normalized drag force in five-dimensional theory with stringy corrections
\be 
F=-v-\frac{1}{2N^2}v^3
\ee 
which implies that our proposal is consistent with the butterfly velocity calculated in \cite{Roberts:2014isa}
\be 
v_B^2=\frac{2}{3}N^2.
\ee 
Finally, let us comment on the relation between drag force and butterfly velocity in quantum systems with spatial anisotropy. Again we expand the dependence of the drag force acting on the charge carrier with the worldline  $x_i=v\cdot t$    as
\be
F_i=-\sigma_i v - {\cal B}_i v^3+...
\ee 
where 
\be 
\sigma_i=\frac{1}{2\pi\alpha^{\prime}} g_{ii}(u_h),\,\,\,\,{\cal B}_i=\frac{1}{2\pi\alpha^{\prime}} \left(\frac{g_{ii}^{\prime}}{g_{tt}^{\prime}}\cdot g_{ii}\right)\Big|_{u=u_h}
\ee 
and using the expression for anisotropic butterfly velocity (see below in the text) we obtain
\be
v_B^{(i)} =\sqrt{\frac{g^{i i}\left(u_{h}\right) g_{t t}^{\prime}\left(u_{h}\right)}{\sum_{k=1}^{d-1} g^{k k}\left(u_{h}\right) g_{k k}^{\prime}\left(u_{h}\right)}}.
\ee 
after some algebra we obtain, that butterfly velocity along the direction $i$ in the presence of anisotropy  is expressed in terms of $\sigma_i$ and ${\cal B}_i$ as \be 
v_B^{(i)}=\sqrt{\frac{\sigma_i^{-1}}{\sum_{i=1}^{d-1}{\cal B}_i\sigma_i^{-2}}}.
\ee 
\subsection{Jet quenching}
\label{app:jq}
It is known \cite{Liu:2006ug} that in some approximation, the jet quenching parameter $\hat q$ admits a non-perturbative formulation as a particular light-like Wilson loop average.  We  consider  the behavior of a Wilson loop stretched between two light-like lines which extend for a distance $L_-$ and in the transverse direction  situated  at a distance $L$ apart. In \cite{Liu:2006ug}, it was shown that the following relation takes place
\be
\left\langle W^{A}(\mathcal{C})\right\rangle \approx \exp ^{-\frac{1}{4 \sqrt{2}} \hat{q} L^{2} L_{-}},
\ee 
when ${\cal C}$ is a rectangle with large extent  $L_-$ in light-like 
 direction and small extent $L$ in a transverse direction.  This relation can serve as a definition of the jet quenching parameter $\hat{q}$.

The calculation of the jet-quenching parameter for anisotropic metric  was presented in  \cite{Giataganas:2012zy}, and we closely follow it.  Again we start with the general form of the metric \eqref{eq:metr-app} and  choose the light-like contour (which corresponds to the direction where partons move) to be oriented along spatial coordinate $x$ 
\be
 x^{\pm}=\frac{1}{\sqrt{2}}\left(t \pm x \right),
\ee 
and the transverse direction corresponding to  the momentum of  radiated gluons to be $y$. In these coordinates the metric has the form
\begin{gather}\nn
d s^{2}=g_{--}\left(d x_{+}^{2}+d x_{-}^{2}\right)+g_{+-} d x_{+} d x_{-}+\\\nonumber+g_{i i(i \neq x)} d x_{i}^{2}+g_{u u} d u^{2}, \\\nn
g_{--}=\frac{1}{2}\left(g_{tt}+g_{xx}\right), \quad g_{+-}=g_{tt}-g_{xx}.
\end{gather}
According to our choice of string configuration we have the ansatz
\be\nn
x_{-}=\tau, \quad y=\sigma, \quad u=u(\sigma),\,\,\,\,
x_{+}, x_{i \neq x} \text { are constant, }
\ee
for which the Nambu-Goto action has the form
\be \nn
S= 2\frac{ L_-}{2 \pi \alpha^{\prime}} \int_{0}^{L/2} d \sigma \sqrt{g_{--}\left(g_{u u} u^{\prime 2}+g_{yy}\right)}.
\ee 
After some algebra we 
obtain that the distance $L$ can be expressed as
\be 
\frac{L}{2}=\int_{0}^{u_{h}} d u \sqrt{\frac{{\cal H}^{2} g_{u u}}{\left(g_{yy} g_{--}-{\cal H}^{2}\right) g_{yy}}},
\ee 
where ${\cal H}$ is defined by
\be 
{\cal H}=\frac{g_{--} g_{yy}}{\sqrt{D}}=g_{--} g_{yy}\Big|_{u^{\prime}=0},
\ee 
and in the last term we evaluate ${\cal H}$ on the string turning point where  $u^\prime=0$. Small ${\cal H}$ corresponds to small distances $L$ and leads us to the expression
\be 
{\cal H}=\frac{L}{2}\left(\int_{0}^{u_{h}} d u \frac{1}{g_{yy}} \sqrt{\frac{g_{u u}}{g_{--}}}\right)^{-1}+\mathcal{O}\left(L^{3}\right).
\ee 
To eliminate the divergences related to  self energy we extract the action of two straight strings
\be
S_{0}=\frac{2 L}{2 \pi \alpha^{\prime}} \int_{0}^{u_{h}} d u \sqrt{g_{--} g_{u u}}.
\ee 
Final expression for the action of the string for small $L$ can be written as
\be 
S=\frac{L_{-} L_{k}^{2}}{8 \pi a^{\prime}}\left(\int_{0}^{u_{h}} d u \frac{1}{g_{k k}} \sqrt{\frac{g_{u u}}{g_{--}}}\right)^{-1},
\ee 
and bringing everything together, we can  obtain that  from the action difference $S-S_0$ for small $L$  the formula for jet quenching parameter has the form
\be 
\hat{q}_y=\frac{\sqrt{2}}{\pi \alpha^{\prime}}{\cal I}^{-1},\,\,\,\,{\cal I}=\int_{0}^{u_{h}} \frac{1}{g_{yy}} \sqrt{\frac{g_{u u}}{g_{--}}}du.
\ee 
where $y$ denotes the direction of momentum broadening.
Now let us calculate the integral ${\cal I}$ for a particular metric considered in the main part of the paper
\be \label{eq:anisgen-app}
ds^2=-\frac{f(z)}{z^{2\nu}}dt^2+\frac{dz^2}{z^{2\nu_z}f(z)}+\frac{dx^2}{z^{2\nu}}+\frac{dy^2}{z^{2\nu_y}}+\sum_{\alpha}\frac{dx_\alpha^2}{z^{2\nu_\alpha}} ,
\ee 
where $f(z)=1-\left(\frac{z}{z_h}\right)^a$. For this metric ${\cal I}$ can be calculated explicitly,  and we are led with the dependence of jet quenching on the horizon location
\be 
q=\frac{\sqrt{2}}{\pi \alpha^{\prime}}\frac{a \Gamma \left(\frac{2 \nu_y-\nu_z+\nu+1}{a}\right)}{\sqrt{2 \pi }  \Gamma \left(\frac{-a+4 \nu_y-2
   \nu_z+2 \nu+2}{2 a}\right)}\cdot z_h^{-2 \nu_y+\nu_z-\nu-1} .
\ee 
\section{Butterfly velocity} \label{sec:butter}
In general, there are two known calculations of butterfly velocity for anisotropic background leading to the same answer.
The first one \cite{Blake:2016wvh,Blake:2017qgd,Jahnke:2017iwi,Fischler:2018kwt} is based on the computation of shockwave on the black hole horizon, and the second one \cite{Gursoy:2020kjd} refers to the subregion duality reconstruction proposed in \cite{Mezei:2016wfz} (for a nice review of recent studies about quantum chaos see \cite{Jahnke:2018off}). In this appendix we briefly review both of these calculations and follow closely to these papers. For a more detailed exposition of each calculation we refer the reader to \cite{Jahnke:2017iwi} and \cite{Gursoy:2020kjd} for each method, respectively.
\subsection*{Shockwave calculation}
Again we start with the metric of the form \eqref{eq:metr-app} and introducing the Kruskal coordinates  $U$ and $V$
\be 
d u_{*}=-\sqrt{\frac{g_{u u}}{g_{t t}}} d u,\,\,\,\,U V=-e^{\frac{4 \pi}{\beta} u_{*}}, \quad \frac{U}{V}=-e^{-\frac{4 \pi}{\beta} t},
\ee 
we rewrite \eqref{eq:metr-app} in the form
\begin{gather} \label{eq:krusk-app}
d s^{2}=2 A(U, V) d U d V+g_{i j}(U, V) d x^{i} d x^{j},\\
A(U, V)=\beta^{2}\frac{g_{t t} (U, V)}{8 \pi^{2}} \frac{1}{U V}.
\end{gather}
In the shockwave setup, we consider the backreaction of the metric \eqref{eq:krusk-app} on the null pulse of energy located at $U=0$ and moving  in the $V-$ direction with the speed of light. The pulse subdivides the spacetime into two regions $L$ and $R$. The metric in the region $R$ remains the same as  \eqref{eq:krusk-app}, whereas the null perturbation modifies the metric in $L$. For the isotropic metric  the back-reaction  can be obtained as a coordinate shift $V \theta(U)\rightarrow V+\alpha\left(t, x^{i}\right)$, where $\alpha\left(t, x^{i}\right)$ is the function to be determined from Einstein equations\footnote{Here $\theta(U)$ is the Heaviside function and $\delta(x)$ is just an ordinary Dirac delta function. Also we redefine $\hat{U}=U$, $\quad \hat{V}=V+\theta(U)$ and $ \hat{x}^{i}=x^{i}$}. For the anisotropic metric following the same logic, we take the ansatz for the shockwave geometry in the form
\be 
d s^{2}=2 \hat{A} d \hat{U} d \hat{V}+\hat{g}_{i j} d \hat{x}^{i} d \hat{x}^{j}-2 \hat{A} \hat{\alpha} \delta(\hat{U}) d \hat{U}^{2},
\ee 
and the shock stress-energy tensor
\be 
T^{\text {shock }}=E e^{2 \pi t / \beta} \delta(\hat{U}) a\left(\hat{x}^{i}\right) d \hat{U}^{2},
\ee 
where $E$ is the energy of the pulse and $a\left(\hat{x}^{i}\right)$ specifies the character of perturbation (for a localized perturbation the choice is $a\left(\hat{x}^{i}\right)=\delta\left(\hat{x}^{i}\right)$).
Following the procedure of solution of Einstein equations  described in \cite{Jahnke:2017iwi} we obtain the condition
\begin{gather} \label{eq:eins-app}
g^{i j}\left(u_{\mathrm{H}}\right)\left[A\left(u_{\mathrm{H}}\right) \partial_{i} \partial_{j}-\frac{u_{\mathrm{H}}}{2} g_{i j}^{\prime}\left(u_{\mathrm{H}}\right)\right] \alpha\left(t, x^{i}\right)=\\=8 \pi G_{\mathrm{N}} E e^{2 \pi t / \beta} a\left(x^{i}\right),
\end{gather} 
defining $\alpha$. The solution corresponding to  $i$-th direction has the form
\begin{gather} 
\alpha(t, z) \sim \exp \left[\frac{2 \pi}{\beta}\left(t-t_{*}\right)-\mu_i \cdot  x_i\right],\\\mu_i^{2}=\left.\frac{u_{\mathrm{H}} }{2 A}g_{ii}\left(\sum_{j=1...d-1} \frac{g_{jj}^{\prime}}{g_{jj}}\right)\right|_{u=u_{\mathrm{H}}}.
\end{gather}
Using this result, we find that  different butterfly velocities $v_B^{(i)}$ associated with the spatial coordinate $x_i$ are given in terms of the coefficient $\mu_i$ 
\be 
v_{\mathrm{B}}^{(i) }=\frac{2 \pi}{\beta \mu_{i}},
\ee 
or explicitly in terms of metric components
\be \label{eq:butterf-anis-app}
v_B^{(i)} =\sqrt{\frac{g^{i i}\left(u_{h}\right) g_{t t}^{\prime}\left(u_{h}\right)}{\sum_{k=1}^{d-1} g^{k k}\left(u_{h}\right) g_{k k}^{\prime}\left(u_{h}\right)}},
\ee 
which for isotropic theory $g_{ij}=g_{xx}\delta_{ij}$ reduces to a single butterfly velocity
\be 
v_B^2 =\frac{ g_{t t}^{\prime}\left(u_{h}\right)}{g_{xx}^{\prime}(u_h)(d-1)}.
\ee 
\subsection*{Entanglement wedge calculation}
An alternative method of  butterfly velocity calculation is due to \cite{Mezei:2016wfz} and we closely follow \cite{Gursoy:2020kjd} for a particular  calculation corresponding to a general anisotropic metric.
Instead of switching to the Kruskal coordinates, we start with the near-horizon expansion of the metric \eqref{eq:metr-app},
which has the form 
\begin{gather} 
d s^{2} \simeq-\left(\frac{2 \pi}{\beta}\right)^{2} \rho^{2} d t^{2}+d \rho^{2}+\\\nonumber +\sum_{i=1}^{d-1}\left[g_{i i}\left(u_{h}\right)+\frac{g_{i i}^{\prime}\left(u_{h}\right)}{g_{t t}^{\prime}\left(u_{h}\right)}\left(\frac{2 \pi}{\beta}\right)^{2} \rho^{2}\right] d x_{i}^{2},
\end{gather}
where Rindler coordinate $\rho$ is defined as 
\be
\left(u_h-u\right)=\left(\frac{2 \pi}{\beta}\right)^{2} \frac{\rho^{2}}{c_{0}},\,\,\,\,c_{0}=-g_{t t}^{\prime}\left(u_h\right).
\ee 
The method of \cite{Mezei:2016wfz} is based on the finding of the smallest entanglement wedge containing the particle at late times. Parametrizing  Ryu-Takayanagi surface $\gamma_{A}$   as $\rho\left(x^{i}\right)$ and fixing the local coordinates $\xi^{i}=x^{i} $,  we obtain the area functional in the near-horizon region
\begin{gather} \label{eq:RT-app}
\operatorname{Area}\left(\gamma_{A}\right)=\sqrt{\operatorname{det} g_{i j}\left(u_h\right)}\times\int d^{d-1} x\Bigg[1+\frac{1}{2}\left(\frac{2 \pi}{\beta}\right)^{2} \times\\\nn \times   \frac{\rho^{2}}{g_{t t}^{\prime}\left(u_h\right)} \sum_{i=1}^{d-1} \frac{g_{i i}^{\prime}\left(u_h\right)}{g_{i i}\left(u_h\right)}+\frac{1}{2} \sum_{i=1}^{d-1} \frac{\left(\partial_{i} \rho\right)^{2}}{g_{i i}\left(u_h\right)}\Big]
\end{gather} 
The equations of motion for  $\rho$ corresponding to this functional have the form
\begin{gather}
\sum_{i=1}^{d-1} g^{i i}\left(u_h\right) \partial_{i}^{2} \rho\left(x^{i}\right)=\mu^{2} \rho\left(x^{i}\right), \\ \mu^{2} \equiv \frac{(2 \pi / \beta)^{2}}{g_{t t}^{\prime}\left(u_h\right)} \sum_{i=1}^{d-1} g^{i i}\left(u_h\right) g_{i i}^{\prime}\left(u_h\right),
\end{gather}
and $\rho$ has the solution in terms of the radius of closest approach to the horizon $\rho_{\text{min}}$ and modified Bessel function of the second kind  
\begin{gather} \nn
\rho\left(\sigma^{i}\right)=\rho_{\min } \frac{\Gamma(n+1)}{2^{-n} \mu^{n}} \frac{I_{n}(\mu|\sigma|)}{|\sigma|^{n}}, \quad \\n = \frac{d-3}{2},\,\,\,\,x^{i}=\sqrt{g^{i i}\left(u_h\right)} \sigma^{i},
\end{gather}
where we also used the rescaled coordinates $\sigma$. We can  estimate the size\footnote{We assume the size of $A$ to be very large.} of the region $A$ denoted by $R_{\sigma}$ in $\sigma^i$ coordinates denoted by     solving the equation given by
\be
\beta \simeq \rho_{\min } \frac{\Gamma(n+1)}{2^{-n} \mu^{n}} \frac{I_{n}\left(\mu R_{\sigma}\right)}{R_{\sigma}^{n}},
\ee
where $R_{\sigma}$ is the size of region $A$ in the coordinates $\sigma^{i} .$ The solution at large $R_{\sigma}$ is:
\be 
\rho_{\min } \simeq e^{-\mu R_{\sigma}}.
\ee 
 Denoting by $R_{i}$  the size of the region $A$ along the $x^{i}$ direction
 \be 
 R_{i}=\sqrt{g^{i i}\left(u_h\right)} R_{\sigma},
 \ee 
  and requiring that $\rho_{\min } \leq \rho(t)$, which is the condition that the infalling particle is contained inside the entanglement wedge, we obtain the relation from \cite{Gursoy:2020kjd} defining the butterfly velocity along each direction
\be 
R_{i} \geq v_B^{(i)} t, \quad  v_B^{(i)} =\sqrt{\frac{g^{i i}\left(u_h\right) g_{t t}^{\prime}\left(u_h\right)}{\sum_{k=1}^{d-1} g^{k k}\left(u_h\right) g_{k k}^{\prime}\left(u_h\right)}}.
\ee